# Theoretical study of the stability of defects in single-walled carbon nanotubes as a function of their distance from the nanotube end

**Feng Ding***

Department of Physics, Göteborg University, SE-412 96, Göteborg, Sweden

## Abstract

Point defects, including atom vacancies, add atom and Stone-Wale defects, close to a (5, 5) single-walled carbon nanotube (SWNT) open end were studied by DFT, semi-empirical PM3 methods and the empirical Brenner potential. It is found that defect stability increases at they become closer to the SWNT open end. Based on these results, a model for removing defects in a growing SWNT is proposed, where the defects diffuse to the SWNT end. Furthermore, the calculations show that the semi-empirical PM3 method compares well with DFT results, and is accurate enough for studying defect formation in SWNTs. In contrast, the empirical Brenner potential yields large errors and is sometimes not even qualitatively correct.

# I. Introduction

As the most important nano-material, carbon nanotubes (CNTs), including single-walled carbon nanotubes (SWNTs) and multi-walled carbon nanotubes (MWNTs), have been studied extensively since their discovery in 1991[1] and 1993[2]. Compared with other nano-materials, such as semiconducting nanowires[3] and nanobelts,[4] CNTs distinguish themselves by their extremely small diameters (e.g., the diameter of the smallest SWNT is 0.4 nm[5]), ultra-long lengths (e.g., isolated SWNTs up to several centimeters [6] and decimeter long SWNT bundles [7] have been produced in experiments), high chemical stability, outstanding mechanical properties (such as tensile strengths of up to 150 GPa, which are more than 300 times that of steel)[8] and excellent electronic properties[9]. All of these favourable physical, chemical and mechanical properties make CNTs the most likely candidate for a diverse range of applications, such as very small electronic devices, ultra-large scale integrated circuits, high density memories, chemical sensors, super-hard and super-strong materials, hydrogen storage and nano-machines.

Although many much progresses have been made to produce CNTs in a controlled way, it is still not sufficiently well controlled to allow for industrial applications. The growth mechanism is not well understood. One important problem that is related to the growth mechanism is how CNT defects are fixed during the nucleation process, especially when CNTs grow at relatively low temperatures.[10, 11] For example, typical temperatures for catalytic chemical vapor deposition (CCVD) CNT growth are 800 K-1500 K,[12] which is several times lower than the melting point of graphite, about 4100 K. Certainly, thermal annealing at these low temperatures cannot fix a high density of defects in condensed carbon structures to form the perfect graphitic layers that are needed to nucleate CNT walls. In the experimental production of CNTs, a lot of amorphous carbon is often formed as the main byproduct.[13] In contrast, carbon atoms near catalyst particles forms high quality graphitic layers,[14, 15] which implies that the catalyst plays a crucial role in healing defects in nucleated carbon structures at low temperatures.

Recent MD simulations also show that, when free carbon atoms incorporate into the open end of a SWNT on a catalyst particle surface, various defects (e.g., pentagons, heptagons, add atoms and vacancies) are readily formed.[16,17,18]. A scooter mechanism has been proposed to explain how a catalyst atom fixes these defects.[10] The catalyst atom stays on the open end of the SWNT and, when a defect is formed, the catalyst atom 'scoot' to the defect position and fix it. If all defects are fixed by the scooter mechanism, the SWNT maintains an open end during the nucleation process. Such a mechanism can explain the healing of defects on the open end of a growing SWNT only. If some defects are not healed at the CNT end during nucleation, they will not be healed by the scooter mechanism and, in the absence of a second healing mechanism, will remain in the CNT wall. That is, if the scooter mechanism were the only way to heal defects, the working efficiency of the catalyst atom must be extremely high to grow pristine SWNTs. For example, the defect density on a high quality SWNT may as low as 1 defect per micron, and if we assume one defect is formed for every 10 added carbon atoms on the tube end then the catalyst atom must fix about 99.99% of the defects. This must be done in a very short time scale since the growth rate of SWNTs is very high (e.g., it has been reported that SWNT grow rates are as large as 20 μm/s in CCVD experiments at 1173 K[19], which means each defect is exposed to the catalyst for just $10^{-5}$ s). Such high catalytic efficiency is not probable. Hence, in order to understand the growth mechanism of high quality CNTs, one should consider more mechanisms for healing the defects, especially for healing the defects in CNT walls.

In this contribution we consider a mechanism where defects in SWNT walls diffuse to the open end of the growing SWNT, where they can be healed by, for example, the scooter mechanism. The position dependence stability of various point defects close to SWNT open end was studied by density functional theory (DFT), the semi-empirical PM3 method[20] and the empirical Brenner potential[21].

## II. Method of the theoretical study

There are three kinds of point defects in CNTs: vacancy, add-atom (AA) and Stone-Wale (SW) defects,[22] which can be formed by removing or adding a carbon atom or rotating a C-C bond by 90 degrees, respectively. All other defects can be obtained by a combination of two or more of these three point defects. Various defects in CNTs are widely observed in experiments.[23, 24]

We chose a section of a (5, 5) SWNT to study the change in stability of defects as a function of their distance from the open end of the nanotube. As shown in panel 0 of Fig. 1, the short SWNT contains 150 C atoms, it is 1.9 nm in length and 20 hydrogen atoms are placed on the dangling bond at both ends. The hydrogen atoms stabilize the nanotube ends to mimic the shape of open ends on catalyst particle surfaces. We create all three types of point defects at different distances from the SWNT open end to study their change in stability as the distance from the end increases.

In the calculations each defect structure was relaxed to its minimum energy using the Brenner potential.[21] Then the semi-empirical PM3 method was used to relax this structure, and then the structure was further optimized using Becke's hybrid three parameter functional[25] and the non-local correlation functional of Lee, Yang, and Parr (B3LYP)[26] DFT method (with 3-21 G basis set). Due to the computational expense of the DFT calculations, only some of the structures were optimized using this method. All of the PM3 and B3LYP calculations were done with the Gaussian98 software package.[27]

## III. Results and discussions

### A. Stone-Wale defects

Stone-Wales (SW) type defects are formed by rotating C-C bonds in the SWNT wall by 90 degrees, which leads to the formation of 5-7-7-5 topological defect (two pentagons connected by a pair of heptagons as shown in panels 2-7 in Fig. 1).[28] SW type defects are believed play a crucial role in the large scale structure transition of different carbon nanostructures, e.g., the coalescence of fullerenes[29] and fusion of carbon nanotubes,[30] and have been studied extensively.

Seven SW defects close to the open end of the SWNT were created by rotating C-C bond 1, 2, 3, 4, 5, 6 and 7 shown in Panel 0 of Fig. 1. As is seen from the figure, the defects shown from Panels 1 to 7 become further away from the SWNT end.

Six structures (Panels 0-5 in Fig. 1) have been calculated using DFT and all of these structures have been studied using PM3 and the Brenner potential. The relative formation energies of these structures obtained by the different methods are shown in Panel 8 of Fig. 1. It is clear that the formation energies obtained by DFT are in good agreement with those obtained by PM3. On the other hand, the formation energies obtained by the Brenner potential are about 40% lower than the DFT results, although the trend is correct. This indicates that the Brenner potential is not sufficiently accurate to give correct quantitative results for defect formation in CNTs. This inaccuracy may be related to the very short C-C bond cut-off length (0.2 nm) in the Brenner potential.

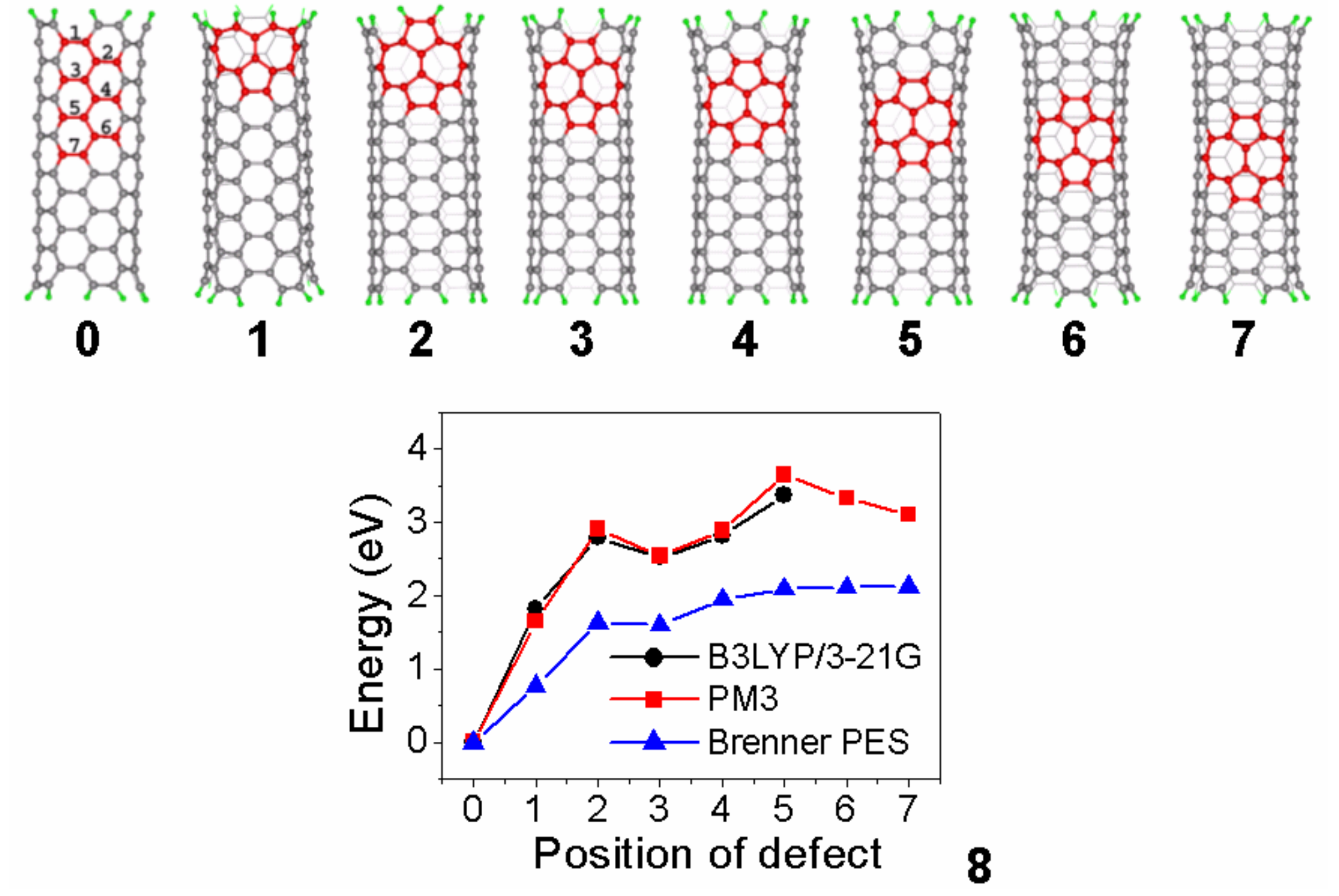

Figure 1. (Color online) Panel 0, a perfect (5, 5) SWNT section; Panels 1-7, the short (5, 5) SWNT with a SW defect at different positions, formed by rotating C-C bonds 1-7 (shown in Panel 0). Panel 8, the relative formation energy of these structures.

Fig. 2 shows three SW structures that are obtained by rotating bond 1 in Panel 0 of Fig. 1, and optimized using B3LYP/3-21G, PM3 and the Brenner potential. The bonds obtained by PM3 are only about 0.6% different in length from those obtained by B3LYP/3-21G, whereas the bond lengths obtained by the Brenner potential differ from the B3LYP/3-21G results by about 3%. In particular, the Brenner potential give an error of 5% for the bond linking two neighboring heptagons.

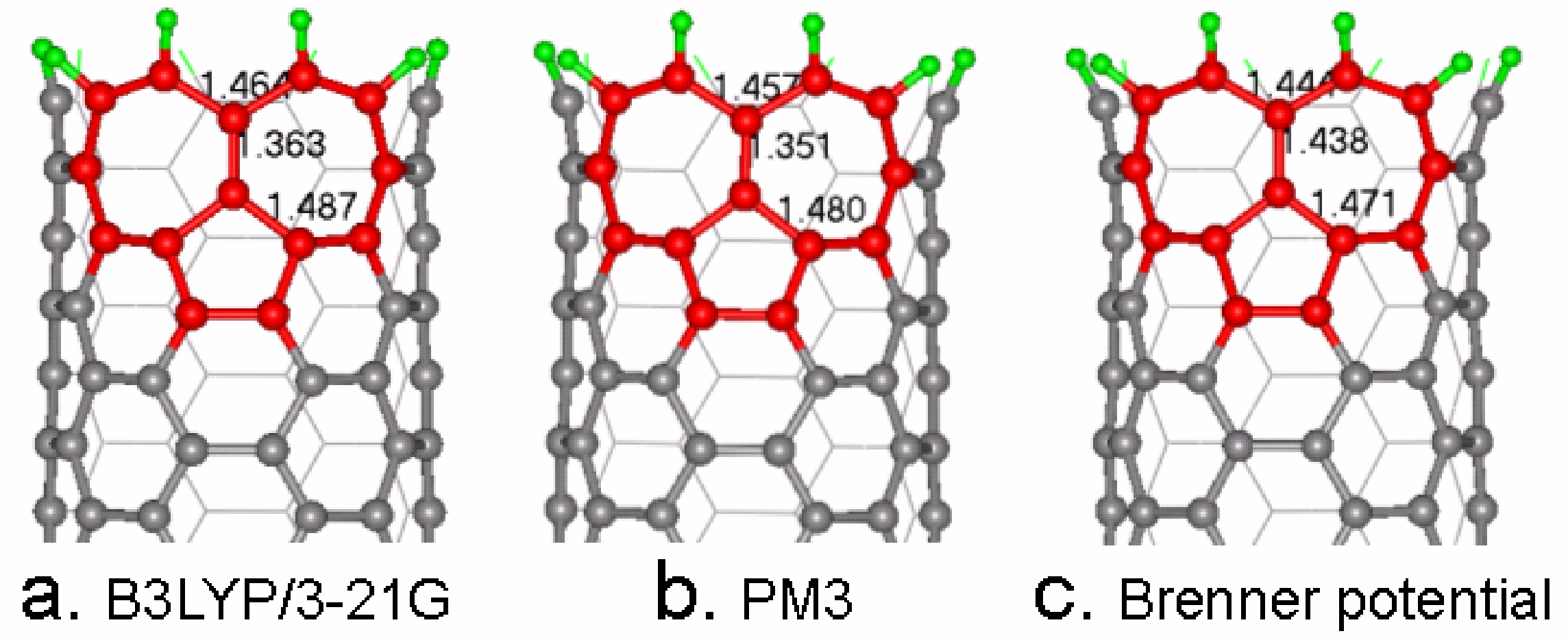

Figure 2. (Color online) Comparison of a SW defect structure (Panel 1 of Fig. 1) optimized by the B3LYP/3-21G, PM3 and Brenner potentials. Note that only part of the SWNT that is used in the calculations is shown in this figure.

As shown in Panel 8 of Fig. 1, the formation energy of these SW defects tend to increase, i.e., they become less stable, as the distance between the defect and the SWNT end increases. For example, the formation increases from 1.65 eV (structure 1) to 2.54 eV (structure 3) and then to 3.09 eV (structure 7).

### B. Add-atom (AA) defects

Add atom defects are also important point defects in CNTs and graphitic layers, which have been studied extensively both theoretically[31],[32] and experimentally[33]. Most recently, the AA defects, together with vacancy defects, induced by electron irradiation have been directly observed by high resolution transition electron microscopy (TEM) by S. Iijima and co-workers.[33]

Recent theoretical study by A. V. Krasheninnikov and co-works indicates that the added carbon may occupy three type positions on an arm-chair SWNT: over a C-C bond perpendicular to the tube axis, over a C-C bond non-perpendicular to the tube axis, inside the SWNT.[32] It is generally found that position over a C-C bond perpendicular to the tube axis is most stable and the position inside the SWNT is most unstable. Especially for (5, 5) SWNT, the absorption energy of a add atom over a perpendicular C-C bond is about 1 and 2 eV higher than that of an add atom over a non-perpendicular C-C bond and inside the SWNT respectively. So, we only the added C atom over a perpendicular C-C bond outside the SWNT is considered in this study (panel a in Fig. 3).

AA defects can easily be created by placing a carbon atom on the SWNT wall over the center of a C-C bond. Eight AA defects corresponding C-C bonds 0-7 as shown in Fig. 3A were considered in this study. Since the study of SW defects shows that the PM3 method is in very good agreement with DFT calculations, the very expensive DFT calculations were not done for these defects.

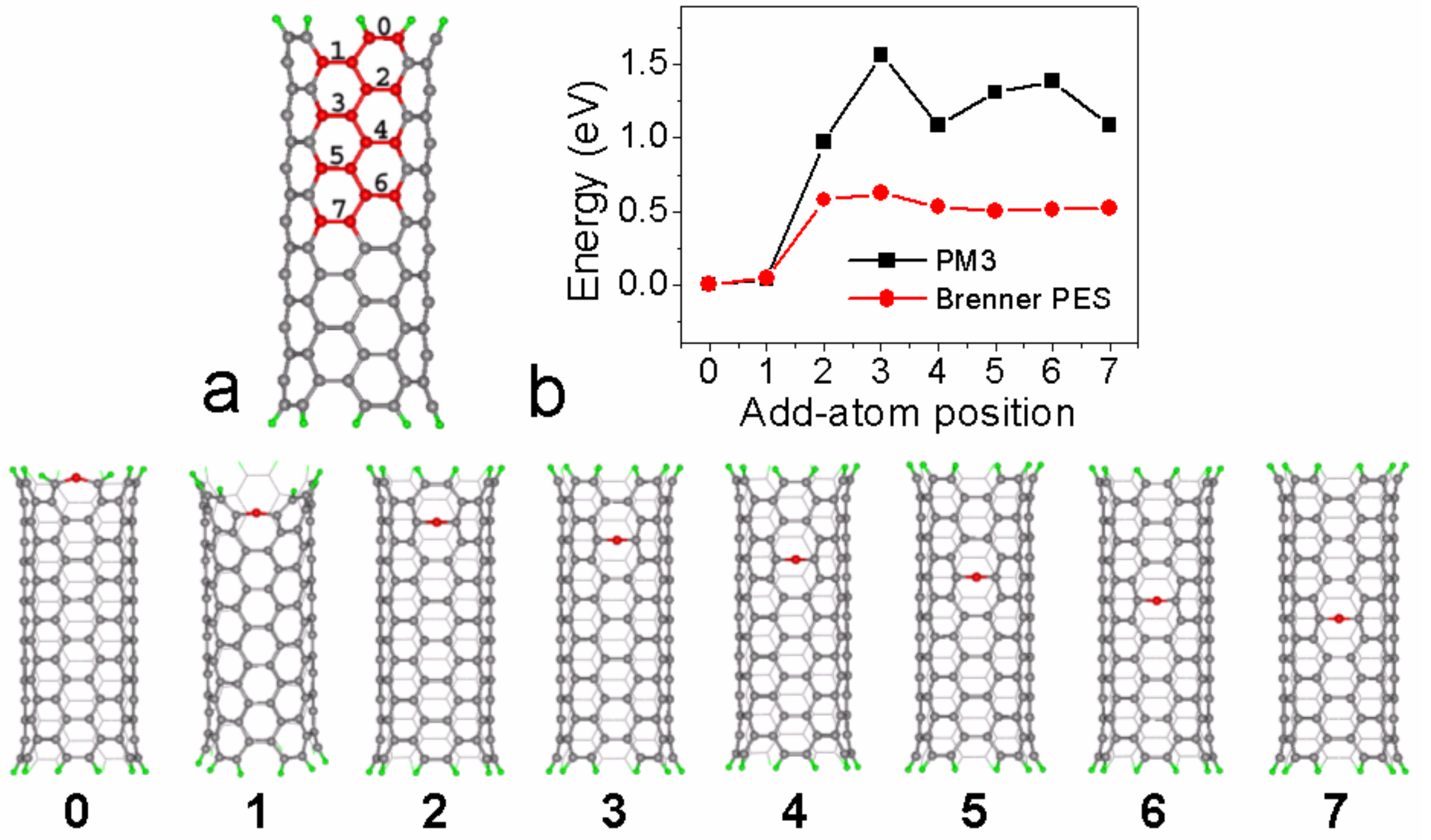

Figure 3. (Color online) The add atom defects and corresponding relative formation energy. 0-7 in Panel A indicate the position where a carbon atom is added. Panel B shows the corresponding relative formation energy of these add atom defects. Panels 0-7 below show the corresponding add atom defect structures optimized by the PM3 method.

Panel B of Fig. 3 shows the relative formation energy of the AA defects. It is clear that the formation energy decreases significantly (about 1.2-1.5 eV) when the AA defect is nearer to the open end of the SWNT. Although the Brenner potential also shows that the add atom is more stable at the open end of the SWNT, the relative formation energies are only 40% of those obtained by PM3.

### C. Vacancy defects

There are numerous studies on vacancy defects in CNT and graphite.[34,35,36,37] It is believed that vacancies in a SWNT can affect its electronic conductance[35] and mechanical properties significantly. [37]

As shown in Fig. 4, one of the 7 red carbon atoms in the (5, 5) SWNT section (0-6 shown in Fig. 4) is removed to form a vacancy defect with differing distances from the end of the SWNT.

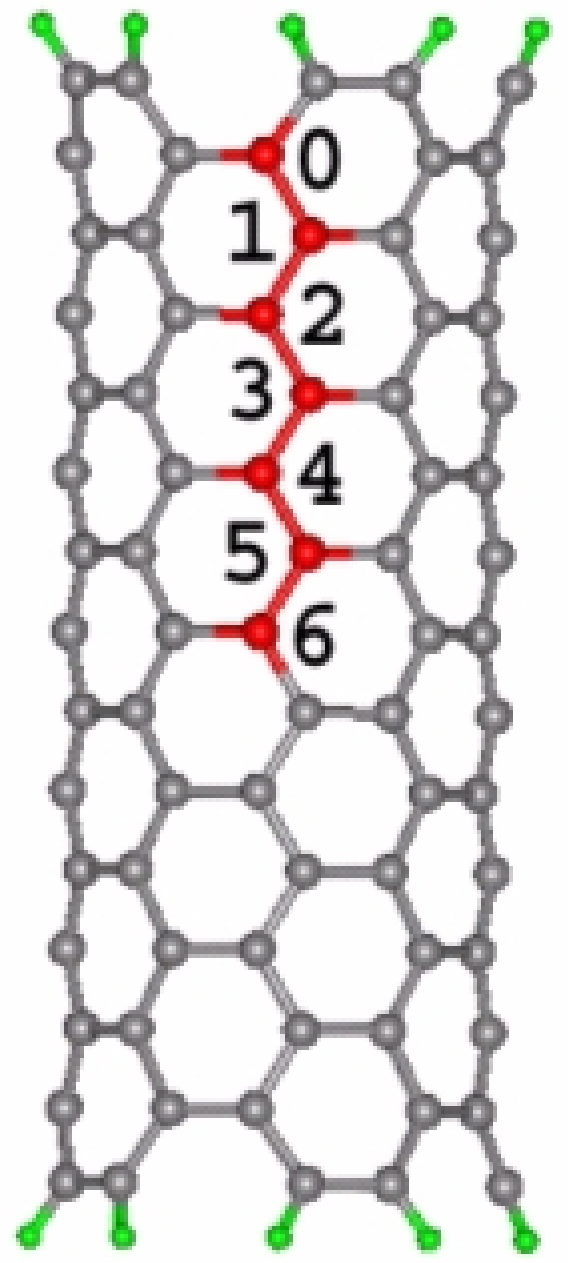

Figure 4. (Color online) The (5, 5) SWNT section used to create vacancy defects by removing one of the 7 carbon atoms identified as 0-6.

Vacancy structures in SWNTs are much more complex than SW and AA defects. A vacancy with three dangling bonds (3DB) is formed at the time when one atom in the nanotube is removed (Panel 0 in Fig. 5). However, 3DB vacancy structure may not be stable. [34, 38] It is found that the 3DB vacancy defect is annealed to one of 6 other possible stable structures. As shown in Panels A-F of Fig. 5, the 6 possible stable vacancy structures can be divided into two classes: A-C are three possible 5-1DB defects (one pentagon and one dangling bond with A having the two upper dangling bonds connected, B the two lower dangling bonds connected and C the upper dangling bond connected to the lower dangling bond) and D-F are three possible structures with one $sp^3$ hybridized carbon atom (with D having the lower dangling bond $sp^3$ hybridized, E having the lower dangling bond $sp^3$ hybridized and F having the intermediate dangling bond $sp^3$ hybridized). The 5-1DB structure is often believed to be the most stable vacancy structure in graphite and SWNTs whereas it is found that the vacancy structure with a $sp^3$ carbon atom is also very stable and, in fact, is sometimes the most stable structure in our study.

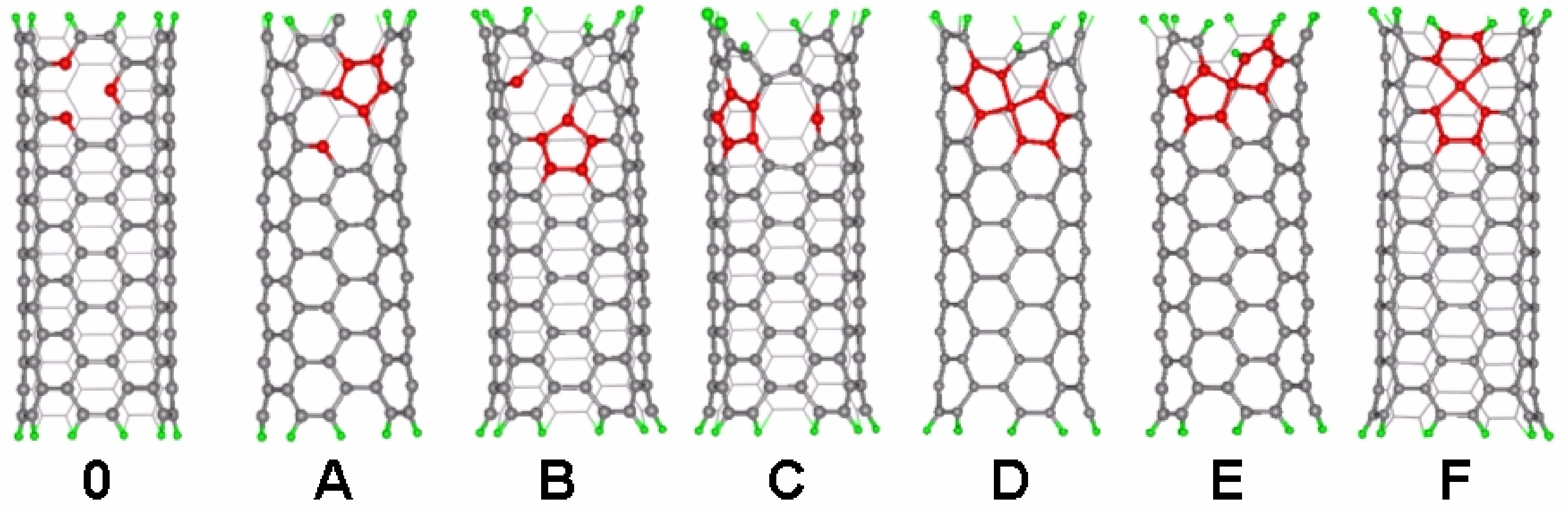

Figure 5. (Color online) Six possible vacancy structures when atom 1 in Fig. 4 is removed. 0 is the unstable 3DB vacancy structure; A-C are 5-1DB type vacancy structures; D-F are $sp^3$ type vacancy structures.

All 6 vacancy structures shown in Fig. 6 may be stable. Hence, a comprehensive study of vacancy formation contains about 42 different structures, which excludes the use of the expensive DFT method. Hence we performed PM3 calculations for all structures and confirmed these results by a limited number of DFT calculations.

The relative formation energy of vacancy structures shown in Panels 1-5 of Fig. 6, obtained from PM3 and DFT, are shown in Panel 6. The differences in formation energies differences obtained from PM3 and DFT calculations are very small, indicating that the PM3 results are quantitatively correct.

The formation energies of these 5 structures obtained from the Brenner potential are also shown in Panel 6 of Fig. 6. The relative formation energy differences obtained from the Brenner potential are only 10-30% of those obtained by the DFT method indicating that, once again, the Brenner potential is not sufficiently accurate for these studies.

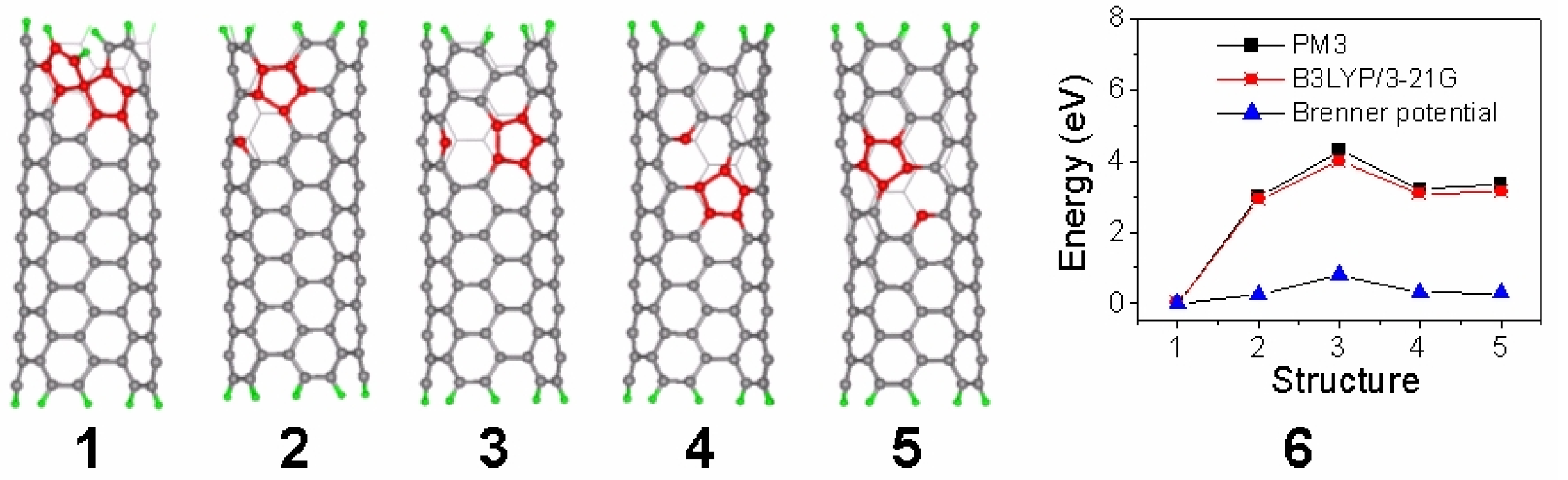

Figure 6, (Color online) Comparison of the relative formation energies of 5 vacancy structures based on PM3, B3LYP/3-21G and the Brenner potential energy surface.

Because of the quantitative accuracy of the PM3 method, as well as it computational efficiency, all 42 possible vacancy structures are studied using this method. These calculations show that many of the possible vacancy structures are unstable and spontaneously anneal to other structures. For example, when carbon atom 1 is removed, all of the three 5-1DB structures are unstable and, when carbon atom 0 is take away, the only stable structure is that with a $sp^3$ hybridized carbon atom and a pentagon as shown in Panel 0 of Fig. 7.

Table I. PM3 energies (in Hartree) of different vacancy structures in the (5, 5). 0-6 in the first line indicates the position of the carbon atom that is removed as shown in Fig. 4. A-F in the first column indicates the sort of vacancy as shown in Fig. 5. Unstable means the structure is not a local minimum.

| | 0 | 1 | 2 | 3 | 4 | 5 | 6 |
|---|---|---|---|---|---|---|---|
| A | 1.66761 | Unstable | 1.85659 | 1.85996 | Unstable | **1.86886** | **1.86486** |
| B | | Unstable | 1.85576 | 1.86239 | 1.86434 | Unstable | 1.86486 |
| C | | Unstable | Unstable | 1.90401 | 1.92510 | 1.92635 | 1.91286 |
| D | | 1.82595 | 1.85576 | 1.86318 | 1.86974 | 1.86972 | 1.86972 |
| E | | **1.74622** | **1.82595** | **1.85576** | **1.86318** | 1.86974 | 1.86972 |
| F | | Unstable | 1.90435 | Unstable | Unstable | Unstable | 1.91640 |
| Minimum | | E | E | E | E | A | A |

The most stable vacancy structures are shown in Panels 0-6 of Fig. 7 and the relative formation energies are shown in Panel 7. Similarly to the results of the SW and AA defects, vacancy defects that are closer to the open end of the SWNT are more stable. Also comparison between the results obtained by the Brenner potential and the PM3 method indicate that the Brenner potential does not provide even qualitatively accurate results for these defects.

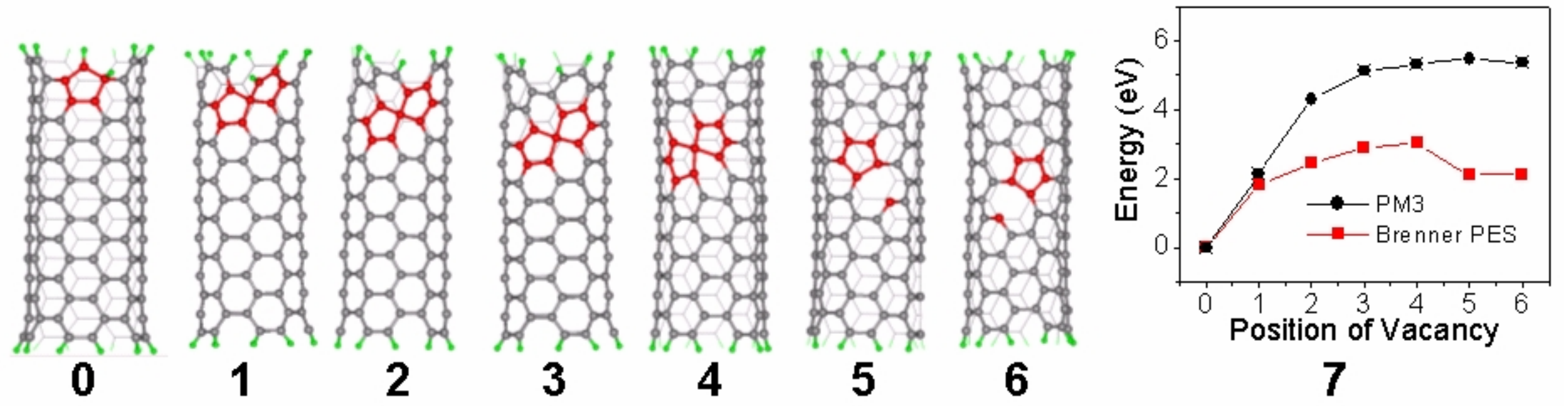

Figure 7. (Color online) Panels 0-6 show the lowest energy vacancy structure formed when removing carbon atom 0-6 in Fig. 4. Panel 7 shows the change in PM3 and Brenner vacancy formation energy at the PM3 level as a function of the distance from the SWNT end.

### D. The mechanism of healing defects during SWNT growth

The generally accepted carbon nanotube growth model is the Vapor-Liquid-Solid (VLS) model.[16, 39,40] According to this model, the nuclear of each SWNT start with a pure supported or free liquid catalyst particle (e.g., Fe, Co, Ni or their alloys). Carbon feedstock (often carbon rich gases, e. g., CO, $CH_4$, $C_2H_2$) decompose on the catalyst particle surface and the released carbon atoms will dissolve to the catalyst particle first. When the catalyst is supersaturated in carbon, the carbon atoms can precipitate to the particle surface and form a fullerene cap and a short SWNT then. With more and more carbon atoms nucleated to the root of the short SWNT, the SWNT can grows longer and longer. Defects are easy formed when carbon atoms add to the SWNT root. How to fix these defects are crucial to understand the growth mechanism of SWNT.

The above study clearly shows that the most stable position of any point defect in a SWNT wall, including SW, AA and vacancies, is at the SWNT's open end. Hence, at elevated temperatures when defects in a SWNT can diffuse along the nanotube wall, they will diffuse from the central part of the SWNT to the open end. There is a catalyst particle at the open end of growing SWNTs. When defects diffuse to the open end the catalyst atoms can heal them via the scooter mechanism. Hence, defects that are not healed during SWNT nucleation, and that remain in the SWNT wall, can still be removed by diffusing to the SWNT end where they are healed by the catalyst. This mechanism is illustrated in Fig. 8.

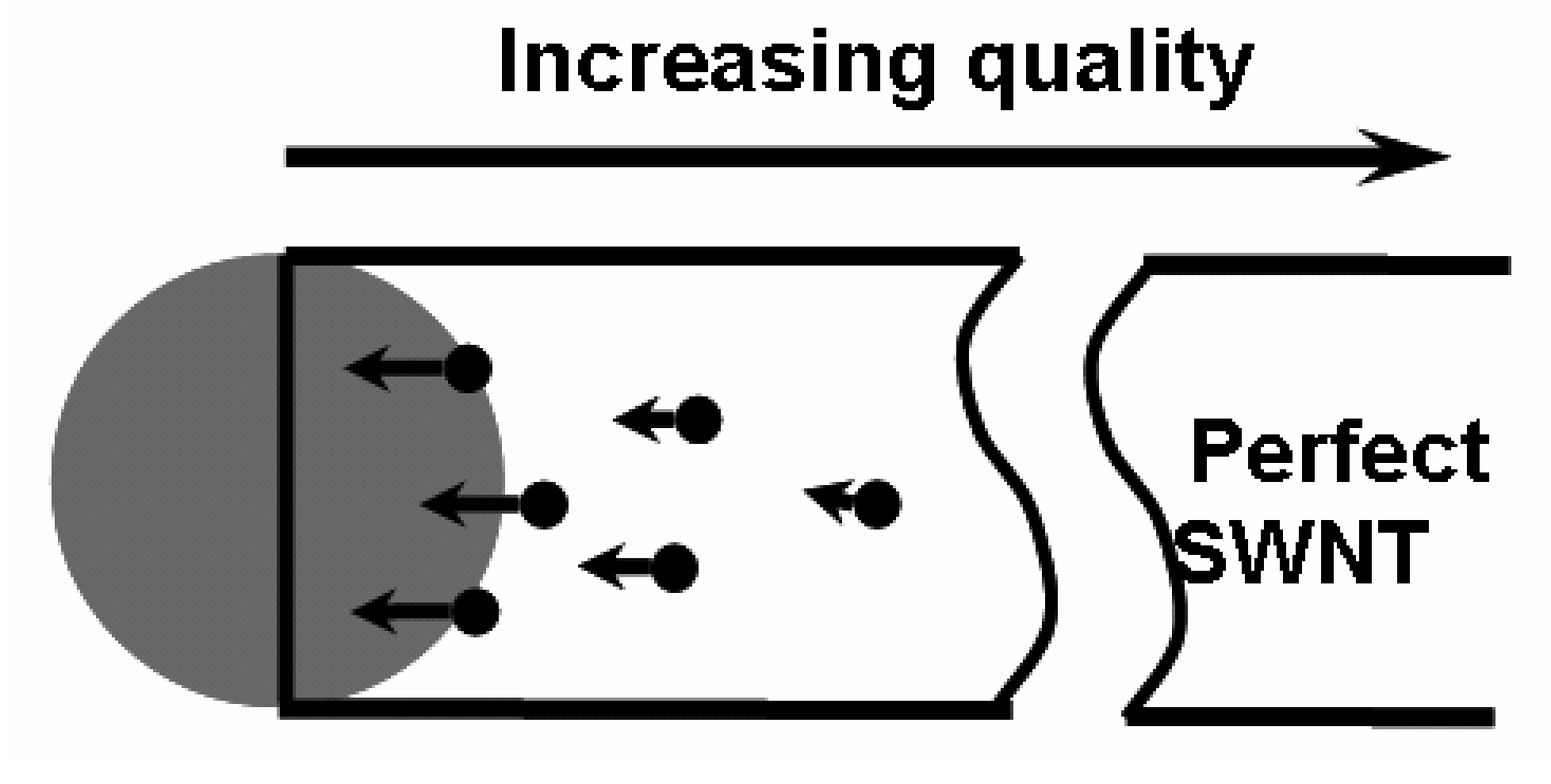

Figure 8. Illustration of the healing of defects in a growing SWNT via diffusion. The black dots are defects and the arrows indicate the diffusion direction. The big sphere

is the catalyst particle.

The above mechanism assumes that defect diffusion along the SWNT wall is rapid. This is supported by many previous experimental and theoretical studies. Certainly if the points defects can diffuse depends on their diffusion path and the corresponding diffusion barrier. The diffusion barrier of a add atom diffusion on armchair SWNT surface is 0.5-1.2 eV, which depends on the tube diameter.[41] Although such a diffusion barrier is little higher than that of add-atom in graphene, 0.47 EV[42], the diffusion of the add carbon atoms is possible at temperature of CNT growth, around 1000-1400 K. The recent reported diffusion barrier of vacancy in graphene is 1.7 eV[36], but such an energy barrier may be highly reduced by the high curvature on the SWNT surface since MD simulations shows that it can diffuse very fast in SWNT.[38] Recent experimental results and MD simulations also show the direct evidence of point defects diffusion in SWNT. Examples include the study by S. Iijima and co-workers, who observed fast diffusion of vacancies, add atoms and defect induced atomic migration in CNT peapods using high resolution tunneling electron microscopy (TEM);[33, 43] P. M. Ajayan et al. have observed surface reconstruction of SWNTs under electron irradiation, and used tight-binding (TB) MD simulations to confirm that this relies on rapid diffusion of vacancies in the SWNT wall at temperatures as low as 700 K.[38] Also, large structural reconstruction of carbon materials, such as the transformation of carbon peapods into double-walled carbon nanotubes (DWNTs) by fusion of encapsulated fullerenes[44, 45] and coalescence of SWNTs[46] is widely observed at temperatures near 1000 $^{o}$C. This is the same as the temperature used for CVD growth of CNTs. The diffusion of SW defects and related topological defect structures in SWNTs and fullerenes is believed to be highly correlated to these transformations.[29, 30] A so-called SW mechanism has been proposed to explain the structural transition caused by rotating C-C bonds in these carbon structures.[29] All of these studies clearly indicate that all of the point defects can diffuse along the SWNT wall at temperatures lower than the melting point of graphite. Clearly, at higher temperatures the diffusion of defects is rapid and they can go from the central part of the SWNT to its open end in a very short time. They are therefore healed more efficiently, and there is a lower density of defects in SWNTs obtained by high temperature production methods. This is seen experimentally since arc discharge and laser ablation produced SWNTs have fewer defects than those produced by CCVD[47]. In addition, CNTs produced in high temperature CCVD often have a higher quality than those produced by low temperature CCVD.[48]

## IV. Conclusion

In conclusion, studies based on DFT, PM3 and the Brenner potential show that the stability of point defects in (5,5) SWNTs increases as they become nearer to the open end of the nanotube. Based on these results, we propose that defects in SWNTs can be healed by their diffusing to the nanotube end, where they are removed by the catalyst particle. It is also found that the PM3 method, which is computationally cheaper than the DFT method, is sufficiently accurate to study defect formation energies and structures. On the other hand, the Brenner potential is only qualitatively correct.

## Acknowledgements

The author is grateful to Prof. Arne Rosén and Dr. Kim Bolton for valuable discussions, as well as for time allocated on the Swedish National Supercomputing facilities and for financial support from the Swedish Foundation for Strategic

Research and the Swedish Research Council.

---

*Electronic address: fengding@fy.chalmers.se

**Captions:**

**Table I.** PM3 energies (in Hartree) of different vacancy structures in the (5, 5). 0-6 in the first line indicates the position of the carbon atom that is removed as shown in Fig. 4. A-F in the first column indicates the sort of vacancy as shown in Fig. 5. Unstable means the structure is not a local minimum.

**Figure 1.** (Color online) Panel 0, a perfect (5, 5) SWNT section; Panels 1-7, the short (5, 5) SWNT with a SW defect at different positions, formed by rotating C-C bonds 1-7 (shown in Panel 0). Panel 8, the relative formation energy of these structures.

**Figure 2.** (Color online) Comparison of a SW defect structure (Panel 1 of Fig. 1) optimized by the B3LYP/3-21G, PM3 and Brenner potentials. Note that only part of the SWNT that is used in the calculations is shown in this figure.

**Figure 3.** (Color online) The add atom defects and corresponding relative formation energy. 0-7 in Panel A indicate the position where a carbon atom is added. Panel B shows the corresponding relative formation energy of these add atom defects. Panels 0-7 below show the corresponding add atom defect structures optimized by the PM3 method.

**Figure 4.** (Color online) The (5, 5) SWNT section used to create vacancy defects by removing one of the 7 carbon atoms identified as 0-6.

**Figure 5.** (Color online) Six possible vacancy structures when atom 1 in Fig. 4 is removed. 0 is the unstable 3DB vacancy structure; A-C are 5-1DB type vacancy structures; D-F are $sp^3$ type vacancy structures.

**Figure 6,** (Color online) Comparison of the relative formation energies of 5 vacancy structures based on PM3, B3LYP/3-21G and the Brenner potential energy surface.

**Figure 7.** (Color online) Panels 0-6 show the lowest energy vacancy structure formed when removing carbon atom 0-6 in Fig. 4. Panel 7 shows the change in PM3 and Brenner vacancy formation energy at the PM3 level as a function of the distance from the SWNT end.

**Figure 8.** Illustration of the healing of defects in a growing SWNT via diffusion. The black dots are defects and the arrows indicate the diffusion direction. The big sphere is the catalyst particle.